\documentclass[conference]{IEEEtran}
\IEEEoverridecommandlockouts
\usepackage{cite}
\usepackage{amsmath,amssymb,amsfonts}
\usepackage{algorithmic}
\usepackage{graphicx}
\usepackage{textcomp}
\usepackage{xcolor}
\usepackage{booktabs}
\usepackage{listings}

\lstdefinestyle{prompt}{
basicstyle=\ttfamily\scriptsize,
columns=fullflexible,
keepspaces=true,
breaklines=true,
breakatwhitespace=false,
postbreak=\mbox{\textcolor{gray}{$\hookrightarrow$}\space},
frame=single,
framesep=3pt,
rulecolor=\color{gray!50},
backgroundcolor=\color{gray!5},
showstringspaces=false,
captionpos=b,
aboveskip=10pt,
belowskip=5pt,
xleftmargin=0pt,
xrightmargin=0pt
}

\usepackage[acronym]{glossaries}
\makeglossaries
\newacronym{llm}{LLM}{Large Language Model}
\newacronym{gpu}{GPU}{Graphics Processing Unit}

\def\BibTeX{{\rm B\kern-.05em{\sc i\kern-.025em b}\kern-.08em
T\kern-.1667em\lower.7ex\hbox{E}\kern-.125emX}}
\begin{document}

    \title{Can Open-Weight LLMs Produce Kernel-Verified Coq Proofs? A Pilot Study}

    \author{
    \IEEEauthorblockN{
    Ahmed Ryan\IEEEauthorrefmark{1},
    Md Erfan\IEEEauthorrefmark{1},
    Akond Ashfaque Ur Rahman\IEEEauthorrefmark{2}, and
    Md Rayhanur Rahman\IEEEauthorrefmark{1}
    }
    \IEEEauthorblockA{
    \IEEEauthorrefmark{1}
    \textit{University of Alabama},
    Tuscaloosa, Alabama, USA\\
    \{aryan9, merfan\}@crimson.ua.edu,
    mrahman87@ua.edu
    }
    \IEEEauthorblockA{
    \IEEEauthorrefmark{2}
    \textit{Auburn University},
    Auburn, Alabama, USA\\
    azr0154@auburn.edu
    }
    }

    \maketitle

    \begin{abstract}
        \glspl{llm} can generate text that resembles a mathematical proof, but resemblance does not establish correctness.
        A formal proof checker verifies correctness by evaluating whether each proof step follows the logical rules.
        Coq bases these rules on the Calculus of Inductive Constructions, a logical framework that defines which proof steps the system may accept.
        This pilot study evaluated 6 open-weight \glspl{llm} on the same 100 theorems from CoqStoq, a benchmark derived from real Coq projects.
        Each \gls{llm} received one attempt per theorem with temperature set to 0, and Coq checked every proposed proof in the theorem's original project environment.
        A proof was counted as successful if Coq's trusted kernel accepted it.

        \textsc{Gemma 4} verified 12 of 100 theorems, \textsc{Llama 3.3} verified 8, and \textsc{DeepSeek Coder V2 Lite} verified 1.
        \textsc{Qwen 3.5}, \textsc{Mistral Small 3.1}, and \textsc{GPT-OSS} verified none.
        The 21 successful proofs covered 15 distinct theorems, and the baseline of standard Coq tactics did not solve 11 of them.
        \glspl{llm} were able to verify theorems that had only short or medium human reference proofs, and no \glspl{llm} were able to verify theorems with long reference proofs.
        Because we explored proof length after selecting the primary analysis, this pattern does not establish that proof length caused the difference.

        For the three \glspl{llm} with at least one success, the total generation cost per verified proof ranged from 741 to 36,193 output tokens, from 14.9 to 178.0 seconds, and from 0.0167 to 0.2000 aggregate GPU hours.
        We did not calculate this ratio for an \gls{llm} that produced no kernel-verified proof.
        Across 600 attempts, the six models produced 21 kernel-verified proofs (3.5\%), which covered 15 distinct theorems.
        The pilot study reports descriptive differences among \glspl{llm} but did not directly test whether the performance difference between any two \glspl{llm} was statistically significant.
        Therefore, the results do not establish a universal ranking of the six models.

    \end{abstract}

    \begin{IEEEkeywords}
        large language models, formal verification, automated theorem proving, Coq, proof generation
    \end{IEEEkeywords}

    \section{Introduction}

    Researchers increasingly use \glspl{llm} to generate formal mathematical proofs.
    A formal proof states each reasoning step in a precise language that a computer program can check.
    Because an \gls{llm} predicts text based on probability rather than enforcing logical validity, its output requires independent verification.
    Coq provides that verification.
    As a proof assistant, Coq helps users state theorems, construct proofs, and check results.
    Coq's small trusted kernel determines whether each proof step follows from the stated assumptions and logical rules.
    The kernel accepts a proof only when every step follows those rules and rejects a proof if it contains an invalid step.

    The difference between a convincing-looking proof and a formally valid proof matters because an \gls{llm} can fail at several levels.
    \glspl{llm} may produce invalid syntax, invoke a tactic that does not exist, refer to a lemma that the project does not provide, or apply a valid theorem under the wrong assumptions.
    A Coq tactic transforms a proof goal into one or more smaller goals.
    An \gls{llm} may also omit a required step, leave a proof incomplete, use definitions that differ from those in the target project, or depend on a different library version.
    These failures show why a proof must be checked rather than judged by how convincing it sounds.
    Therefore, we count a proof as successful only when Coq's kernel accepts it in the theorem's original project environment.

    Existing studies do not yet establish how reliably general-purpose open-weight \glspl{llm} produce verified proofs for theorems collected from real Coq projects.
    The literature also provides limited evidence about three related matters: which sampled theorems these \glspl{llm} solve, whether their successes overlap with basic Coq automation, and how many computational resources the \glspl{llm} consume.
    Coq automation uses tactics to search for a proof with limited or no step-by-step direction from a user.
    We therefore use one fixed experimental procedure to evaluate proof success, compare model and automation results, and measure resource use.

    \textit{The goal of this study is to evaluate how well open-weight \glspl{llm} produce kernel-verified Coq proofs, identify the theorems they prove and measure the resources they use, and eventually provide empirical evidence of the capabilities of \glspl{llm} to the researchers, tool developers, practitioners, and educators.}
    To achieve this goal, we investigate the following research questions:

    \begin{itemize}

    \item{\textbf{\textit{RQ1}}: \textit{How often do \glspl{llm} generate a Coq proof that the kernel accepts?} }

    \item{\textbf{\textit{RQ2}}: \textit{Which sampled theorems do \glspl{llm} solve, and how do their success compare with basic Coq automation?} }

    \item{\textbf{\textit{RQ3}}: \textit{How much computational resource do \glspl{llm} use while generating verified proofs?} }

    \end{itemize}




    For $RQ_1$, each of six \glspl{llm} attempts 100 theorems from real Coq projects, and we count only proofs accepted by the Coq kernel.
    We report verification rates with confidence intervals to express statistical uncertainty.
    For $RQ_2$, we compare verified theorems across projects, reference-proof lengths, and a fixed Coq-automation baseline.
    This comparison shows whether the models and baseline prove the same or different theorems.
    For $RQ_3$, we measure generated tokens, wall-clock time, and GPU time.
    Because every model attempts each theorem once, these measurements support a consistent comparison of resource use.

    \vspace{5pt}

    \textbf{\textit{Contributions}}: We list our contributions as follows:

    \begin{itemize}

    \item{A reproducible benchmarking protocol that runs each model in the theorem's original Coq environment and accepts only proofs verified by the Coq kernel.}

    \item{An evaluation of six open-weight \glspl{llm} under the same one-attempt proof-generation rule.} 

    \item{A comparison of the theorems solved by each \gls{llm} and a fixed Coq-automation baseline.}

    \item{A resource analysis based on generated tokens, elapsed time, and GPU time.}

    \end{itemize}    






    \section{Key Concepts}

    We use the following concepts to define the study and interpret its reported outcomes.

    \textit{Coq} checks whether a proof follows its formal rules, which are based on a logical system called the Calculus of Inductive Constructions.

    A \textit{formal proof} expresses each reasoning step using defined symbols, syntax, and logical rules that a computer program can interpret and verify.

    A \textit{theorem} states a claim that follows from previously defined assumptions, definitions, and logical rules.
    In this study, each theorem specifies the claim for which an \gls{llm} must produce a proof that Coq can verify.

    A \textit{proof goal} states what needs to be proved at a particular stage of the proof.
    As the \gls{llm} applies valid proof steps, Coq updates the goal until no unproved goals remain.

    A \textit{tactic script} directs Coq through a sequence of commands that reduce a proof goal to smaller goals until no goal remains.
    Even after the script closes every goal, the complete proof must pass the kernel checks.

    An \textit{large language model (LLM)} is a machine-learning model trained on large datasets to generate new content from a given prompt.
    In this study, an \gls{llm} receives a Coq theorem statement and the permitted project context, then generates a tactic script intended to prove the theorem.

    \textit{CoqStoq} is a dataset of Coq theorems collected from real software projects.
    It provides each theorem with its project context, including the definitions, assumptions, libraries, and dependencies available when the theorem was written.

    A \textit{project environment} provides the notation, earlier results, tactics, library paths, and software configuration that Coq requires to compile a theorem.
    We verify each \gls{llm}-generated proof in its theorem's original project environment because each project uses different settings.

    A \textit{Reference Proof} is the human-written solution provided by CoqStoq.
    We use this proof to confirm that each theorem has a valid solution and to group the theorems by proof length.
    The \glspl{llm} do not receive or see the reference proofs.

    \textit{verified@1} indicates whether a \gls{llm}’s first and only proof attempt for a theorem passes all verification checks.
    It measures success from a single attempt and does not represent performance when a \gls{llm} can generate multiple proofs or revise a failed proof.

    \section{Related Work}

    \subsubsection*{Coq Proof-Generation Systems}

    Several systems learn to generate Coq proofs.
    CoqGym collects proofs from existing projects and supports ASTactic, which learns to generate tactics from those proofs~\cite{yang2019learning}.
    Proverbot9001 also learns to generate Coq proofs and evaluates its method mainly on CompCert, a project included in our sample~\cite{sanchezstern2020generating}.
    Tactician learns from existing tactic scripts and can suggest the next tactic or attempt to complete a proof~\cite{blaauwbroek2020tactician}.
    Rango provides the closest comparison because it also evaluates proof generation on CoqStoq~\cite{thompson2025rango}.
    However, Rango uses a fine-tuned model and retrieves relevant information before generation.
    In contrast, our pilot evaluates general-purpose chat models with one fixed prompt, one attempt, and no fine-tuning, retrieval, repair, or interactive feedback.

    \subsubsection*{Benchmarks and Automation Baselines}

    Benchmark studies show why an overall success rate is not sufficient.
    CoqStoq contains theorems from real Coq projects and preserves differences in libraries, notation, and proof styles~\cite{thompson2025rango}.
    By comparison, miniF2F contains competition-style mathematical problems for several proof assistants~\cite{zheng2022minif2f}, and LeanDojo supports proof search in Lean projects~\cite{yang2023leandojo}.
    Because these benchmarks use different theorem sources and proof assistants, their success rates cannot be compared directly.
    Existing Coq automation provides another reference point.
    CoqHammer uses relevant prior results and external theorem provers to construct proofs that Coq can verify~\cite{czajka2018hammer}.
    Our fixed baseline of \texttt{auto}, \texttt{eauto}, and \texttt{intuition auto} is more limited but easier to inspect and apply consistently.
    Thus, an LLM-only success means that this fixed baseline failed, not that all Coq automation would fail.

    \subsubsection*{Computational Resource Measurement}

    Prior systems use different models, search procedures, hardware, benchmarks, and stopping rules, which prevents direct comparisons of computational cost.
    For example, systems that retrieve examples, try multiple tactics, repair failures, or interact repeatedly with Coq perform more work than a one-attempt system.
    We therefore report generated tokens, generation time, and GPU time only for our protocol and hardware.
    For each model with at least one verified proof, we also divide its total resource use by its number of verified proofs.
    This calculation includes failed attempts because they contribute to the total cost of obtaining the successful proofs.

    \subsubsection*{Position of This Study}

    This paper presents a descriptive pilot evaluation of six general-purpose open-weight LLMs on the same 100 CoqStoq theorems.
    Unlike Coq-specific systems, it measures proof generation without task-specific training, retrieval, repair, repeated attempts, or interactive feedback.
    Kernel verification addresses $RQ_1$.
    Analyses by project, reference-proof length, and automation overlap address $RQ_2$.
    Resource measurements address $RQ_3$.
    The study therefore provides an initial account of one-attempt proof generation under a fixed protocol rather than a universal ranking of models or a direct comparison with systems that use different forms of assistance.

    \section{Methodology}

    \begin{figure*}[t]
        \centering
        \includegraphics[width=\textwidth]{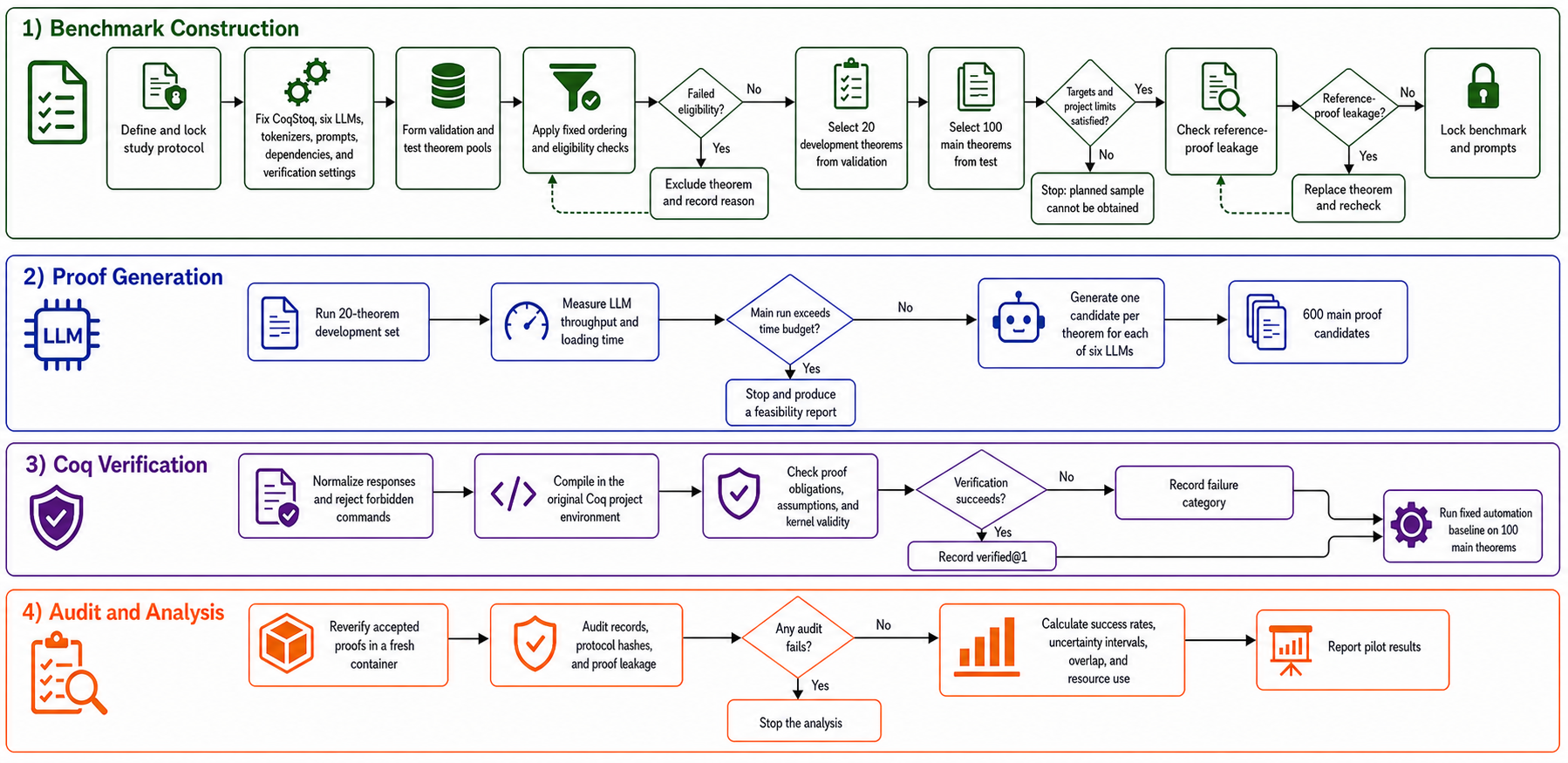}
        \caption{Methodology of the pilot study.}
        \label{fig:methodology}
    \end{figure*}
    
    We conduct the pilot study in four phases.
    First, we fix the study settings and construct the theorem sets.
    Second, we prompt the six \glspl{llm} and collect their proposed proofs.
    Third, Coq verifies every proposed proof, and we compare the verified results with a fixed automation baseline.
    Finally, we audit the complete record and calculate the reported statistics.
    Because we fix the benchmark, prompts, and verification rules before observing the main results, those results cannot influence the evaluation rules.

    \subsection{Study design and benchmark construction}

    \subsubsection{Fix the protocol and software versions}

    Before generating any study proof, we record every setting that can affect generation or verification.
    We record theorem-selection rules, model versions, prompts, software dependencies, generation options, and verification rules because changing any item may change the outcome.
    We clone the official CoqStoq repository and its included dependencies.
    We then fix the copy to one Git commit SHA, which uniquely identifies an exact repository version.
    We evaluate six \glspl{llm} shown in Table~\ref{tab:model-details}.
    We fix each LLM repository and its tokenizer to an unchangeable Hugging Face commit SHA.
    We obtain each identifier through the Application Programming Interface (API) of Hugging Face.

    We compute cryptographic hashes for three types of content: the prompt template, the complete prompt sent to each \gls{llm} after formatting, and the theorem context included in that prompt.
    A cryptographic hash acts as a digital fingerprint because any change to the content changes the hash.
    We record these hashes with the code commit, container-image digests, dependency-lock hashes, and all generation and
    verification settings.
    Before generating proofs, the system checks that the current LLM versions, prompts, dependencies, verification environment, and settings match the recorded configuration.
    These checks prevent unrecorded changes from affecting the results and allows reproducibility of the experiment.

    \subsubsection{\gls{llm} Selection}

    We select the six \gls{llm} using four criteria.
    First, each \gls{llm} provides downloadable weights and supports local execution.
    Second, each \gls{llm} can follow instructions and generate source code without task-specific fine-tuning.
    Third, each \gls{llm} fits within our available computational resources (192GB GPU Memory).
    Fourth, the selection includes \glspl{llm} from different developers and parameter scales to represent a range of open-weight model families.
    We also include \textsc{DeepSeek Coder V2 Lite Instruct} as a code-specialized model to examine whether training focused on source code benefits Coq proof generation.
    These criteria yield the six \glspl{llm} in Table~\ref{tab:model-details}.

    \begin{table}[htbp]
        \vspace{-10pt}
        \centering
        \scriptsize
        \setlength{\tabcolsep}{5pt}
        \caption{Language models evaluated in the study}
        \label{tab:model-details}
        \begin{tabular}{lllcc}
            \toprule
            \textbf{Label} &
            \textbf{Full Model Name} &
            \textbf{Developer} &
            \textbf{Params.} &
            \textbf{CW} \\
            \midrule

            Llama
            & Meta Llama 3.3 70B Instruct
            & Meta
            & 70B
            & 128K \\

            Qwen
            & Qwen 3.5 35B-A3B
            & Alibaba/Qwen
            & 35B
            & 262K \\

            Gemma
            & Gemma 4 31B IT
            & Google
            & 31B
            & 256K \\

            Mistral
            & Mistral Small 3.1 24B Instruct
            & Mistral AI
            & 24B
            & 128K \\

            GPT-OSS
            & GPT-OSS 20B
            & OpenAI
            & 20B
            & 128K \\

            DeepSeek
            & DeepSeek Coder V2 Lite Instruct
            & DeepSeek
            & 16B
            & 128K \\

            \bottomrule
        \end{tabular}
        \\[3pt]
        \raggedright\scriptsize
        Note: Params. denotes the nominal total parameter count reported for each
        model; B = billion parameters; CW = context window; K = thousand tokens.
        \vspace{-5pt}
    \end{table}

    \subsubsection{Construct the development and main sets}

    We use CoqStoq as the source repository.
    CoqStoq divides its theorems into predefined validation and test splits.
    We select 20 theorems from the validation split to form the development set, which we use to test the methodology.
    We select 100 theorems from the test split to form the main set, which we use to produce the reported results.
    Each theorem belongs to a Coq project.
    Each project defines notation, reusable results, automated tactics, and library paths.
    Together, these components form the theorem's project environment.
    Because a proof may work in one project and fail in another, we compile and verify every proof in its original project environment.
    This procedure verifies each proof with the project-specific settings that Coq requires for that theorem.

    We treat every theorem in the CoqStoq test split as a \textit{candidate theorem}, meaning that it may be selected for the main evaluation set if it meets the selection requirements.
    We assign each candidate a reproducible position based on its split, a fixed study date, and its unique identifier.
    We then sort and examine the candidates in that order.
    This procedure prevents manual selection and produces the same ordering when researchers repeat the study.
    We include a candidate theorem only if it satisfies three conditions:
    (a) The theorem's project contributes fewer than 10 selected theorems.
    (b) The theorem's statement differs from every selected statement.
    (c) The theorem passes the eligibility checks described below.

    To determine eligibility, we test each theorem in a container that provides an isolated environment with fixed tools and dependencies.
    We first build the theorem’s project.
    We then place its original human-written reference proof in a separate copy of the source file and ask Coq to verify it.
    Using a copy prevents the check from changing the original project.
    We exclude a theorem if its project cannot be built or Coq cannot verify its reference proof.
    We use the reference proof only for this eligibility check and the later proof-length analysis, we never show it to an LLM.

    Because this is a pilot study, we limit each prompt to 24,000 tokens to balance the amount of project context with the required memory, runtime, and GPU resources.
    We exclude a theorem if its prompt exceeds this limit under any model’s tokenizer.
    Applying the same limit across all models keeps the evaluation set identical and leaves space for proof generation.
    We also exclude a theorem if its prompt contains its own reference proof or another selected theorem’s reference proof, which prevents models from copying known solutions.
    We continue the selection process until the main set contains exactly 100 theorems from at least 10 projects.
    We select no more than 10 theorems from any project to limit the influence of large projects and represent a wider range of coding styles and library environments.

    We apply the same procedure to CoqStoq’s validation split to select 20 development theorems.
    We exclude projects represented in the main set so that the development and main sets contain different projects.
    The development and main sets contain different projects to prevent test-set leakage.
    This separation ensures that adjustments made during development are not tailored to the projects used for the final evaluation.
    We record why we exclude each candidate theorem.
    If the fixed selection rules cannot produce both sets, we stop the experiment rather than change the rules after examining the candidates.

    \subsubsection{Prevent reference-proof leakage}

    We show each \gls{llm} only the part of the source file that appears before the selected theorem.
    We call this material the source prefix.
    Because the source prefix ends where the theorem begins, it excludes that theorem's reference proof and all later file content.

    Two selected theorems may come from the same file.
    The proof of the earlier theorem may then appear in the later theorem's source prefix.
    We call this event cross-theorem leakage, which means that one evaluated theorem's reference proof becomes visible in another evaluated theorem's prompt.
    We remove the affected theorem, replace it with the next eligible theorem in the fixed order, and repeat the check until no leakage remains.

    Before locking the protocol, we audit every final prompt for reference-proof exposure.
    We refuse to lock the protocol if the audit finds any exposure.
    These checks remove known routes by which an \gls{llm} could copy a selected reference proof from the prompt.

    \subsection{Prompting and Proof Generation}

    \subsubsection{Apply one prompt design to every LLM}

    We use the same fixed prompt template for every \gls{llm} and theorem (see Listing~\ref{lst:prompt-template}).
    For each theorem, we replace only \texttt{source\_prefix} and \texttt{theorem\_statement} with the relevant project context and theorem statement.
    We provide no model-specific examples, hints, retrieval results, repair attempts, or other assistance.

    \begin{lstlisting}[
        style=prompt,
        caption={Fixed prompt template used for all models and theorems.},
        label={lst:prompt-template}
        ]
 Complete the following Coq theorem. Return only the proof
 body: the commands that belong between `Proof.` and `Qed.`
 Do not include Markdown fences, `Proof.`, `Qed.`,
 admissions, new axioms, or commentary.

 Project-visible source: {{ source_prefix }}

 Theorem: {{ theorem_statement }}\end{lstlisting}

        We then format the resulting message with each model's official chat template.
        This formatting changes the model-specific control tokens and markup but not the instructions or theorem content.
        The prompt also tells the LLM to omit Markdown fences, proof markers, unfinished-proof commands, and new assumptions.
        These restrictions standardize the responses and prevent the LLM from bypassing the intended task.
        We record hashes of the prompts to confirm that the protocol remains fixed.

        \subsubsection{Generate one candidate for each theorem by each \gls{llm}}

        Each LLM generates one candidate proof for each of the 100 main theorems and 20 development theorems.
        Therefore, The six \glspl{llm} produce 600 main candidates and 120 development theorem.
        Each candidate proof lets us measure `verified@1`, which records whether an \gls{llm}'s first and only candidate passes verification.
        This design controls computational cost, but it does not estimate success under repeated sampling.

        We set the generation temperature to 0.
        When generating a response, an \gls{llm} considers several possible tokens for the next position and assigns each token a probability based on the prompt and the tokens already generated.
        A temperature of 0 directs the \gls{llm} to select the token with the highest estimated probability instead of randomly sampling from the alternatives.
        This setting reduces variation and makes repeated runs more consistent, although differences in software or hardware may still affect exact reproducibility.
        We also limit each response to 4,096 generated tokens.
        This limit provides enough space for a substantial Coq proof while preventing unusually long responses from consuming excessive runtime, memory, or GPU resources.

        We distribute each \gls{llm}'s fixed parameters across four \glspl{gpu}.
        Before timing the study attempts, we run one non-study prompt as a warm-up.
        A warm-up performs an initial generation so that one-time setup work does not inflate the measured time for the first theorem.
        For each attempt, we record the prompt hash, raw response, input and output token counts, stopping reason, timestamps, and retry information.
        We run the six \glspl{llm} sequentially.
        Each \gls{llm} completes the development set before any \gls{llm} begins the main set.
        We unload one \gls{llm} and confirm that the \glspl{gpu} have released its memory before we load the next \gls{llm}.
        This sequence prevents the \glspl{llm} from competing for the same hardware and keeps timing conditions more consistent.

        \subsubsection{Confirm that the main run fits the time budget}

        During the development run, we measure each \gls{llm}'s throughput and loading overhead.
        Throughput measures how many theorems an \gls{llm} processes per unit of time.
        Loading overhead measures the time required to prepare the \gls{llm} before generation.
        We use only these timing measurements to estimate the duration of the 600 main generations.
        We do not examine development-set success when making this decision, because early success results must not influence the final sample.

        We start the main run only if the estimate fits within 40 hours of the 48-hour pilot budget.
        We reserve the remaining 8 hours for repeat verification, auditing, analysis, and reporting.
        If the estimate exceeds 40 hours, we stop and produce a feasibility report.
        We do not reduce the fixed sample to fit the budget, because that change would alter the evaluation plan.

        \subsection{Verification and Automation Baseline}

        \subsubsection{Verify every candidate with Coq}

        We first normalize each \gls{llm} response.
        Normalization converts a small set of recognized output formats into the proof-body format expected by the verifier.
        We remove at most one surrounding Markdown code fence and reject any wrapper that we cannot interpret without guessing.
        These changes standardize presentation without changing the proposed proof.
        We insert the normalized proof body between fixed `Proof.` and `Qed.` commands in a copy of the theorem's source file.
        We then apply a syntax-aware scan, which follows Coq's command structure instead of treating every period as a command boundary.
        This approach avoids misreading comments, notation, and qualified names.
        The scan rejects commands that could bypass the task: `Admitted`, `admit`, `Axiom`, `Conjecture`, `Parameter`, `Abort`, and `Declare ML Module`.

        Next, we compile the file with Coq 8.18 in the theorem's original project directory.
        We use the project's library-path mappings, which connect names in the proof to the required files and modules.
        We require Coq to finish compilation and leave no open proof obligations.
        An open proof obligation is a remaining goal that still requires a proof.
        We then run `Print Assumptions`, which lists every assumption used by the completed theorem.
        We reject any assumption that the protocol does not allow.
        Finally, we run `coqchk`, which asks the Coq kernel to check the compiled library.
        Together, these checks detect incomplete proofs, added assumptions, and proofs verified under the wrong project setup.

        We classify a candidate as verified only if it compiles, closes every proof obligation, uses
        no forbidden assumption, and passes `coqchk`.
        The `verified@1` measure reports whether the one candidate generated for a theorem satisfies all four conditions.
        Thus, every reported success represents a complete proof accepted by Coq rather than a script that only appears plausible.
        We assign each failed attempt one predefined category, such as `syntax\_error`, `tactic\_failure`, `unauthorized\_assumption`, or `infrastructure\_error`.
        An infrastructure error describes a failure in the evaluation system rather than a failure in the proposed proof.
        We retry only infrastructure errors, and we never change the candidate during a retry.
        This rule prevents a verification problem from giving an LLM an additional proof attempt.

        \subsubsection{Compare the LLMs with a fixed automation baseline}

        We also evaluate a fixed automation baseline.
        An automation baseline is a consistent procedure that provides a reference point for interpreting \gls{llm} performance.
        We apply the following \gls{llm}-independent tactic script to all 100 main theorems:

        \begin{lstlisting}[
            style=prompt,
            caption={Automation Baseline Tactic Script},
            label={lst:automation-baseline}
            ]
first [ solve [ auto ] | solve [ eauto ] | solve [ intuition auto ] ].\end{lstlisting}

            The script asks three standard Coq tactics, in order, to solve the complete theorem.
            We run the script with the same timeout, assumption rules, project environment, and verification checks
            used for the LLM candidates.
            This comparison identifies LLM successes that overlap with basic built-in proof search.
            It does not represent the strongest automation available for Coq.

            \subsection{Outcomes, Statistical Analysis, and Reproducibility}

            \subsubsection{Reverify and audit every result}

            We check every accepted proof again inside a fresh verification container that did not perform the first check.
            Repeating verification in a newly created environment helps detect results that depended on stale files, cached build products, or accidental state from an earlier run.
            Before calculating any statistic, we confirm all of the following:
            (a) the benchmark contains exactly 100 main theorems and 20 development theorems.
            (b) the records name exactly six fixed model versions.
            (c) all 600 model-theorem combinations appear exactly once.
            (d) every generation has a final result that is not an infrastructure error.
            (e) every accepted proof again passes the assumption and coqchk checks.
            (f) every prompt and protocol hash matches the locked value, and
            (g) the final leakage audit reports no reference-proof exposure.
            The analysis stops if any check fails.
            Requiring a complete and internally consistent record prevents missing, duplicated, or unresolved attempts from silently changing a model's reported rate.

            \subsubsection{Calculate statistics only from audited records}

            For each model, we report its verified@1 count out of 100 theorems.
            We present the theorem-level success rate with a Wilson 95\% confidence interval.
            We also report the average success rate across projects with a project-clustered bootstrap 95\% interval based on 10,000 resamples and seed 20260728.
            The bootstrap resamples whole projects because theorems from the same project may be related.

            The theorem-level rate shows the percentage of all sampled theorems that a model verifies, whereas the project-level rate gives each project equal weight.
            We treat the project-level result as secondary because the sample contains only 12 projects.
            For models with no successes, the bootstrap interval is [0\%, 0\%] because every resampled project also contains no successes.
            This result does not prove that the model’s broader success probability is zero.

            We also report results by project, reference-proof length, overlap with the automation baseline, and computational cost.
            We treat proof-length results as exploratory because reference-proof length only roughly represents difficulty.
            Finally, we report results only after all 600 attempts produce valid outcomes, ensuring that every model retains a denominator of 100 and that infrastructure failures are not counted as proof failures.

            \subsubsection{Reproducibility and analysis deviations}

            The results cover verification rates and confidence intervals ($RQ_1$), performance by project, reference-proof length, and automation overlap ($RQ_2$), and computational cost ($RQ_3$).
            We classify theorems using their human-written reference proofs: short proofs contain at most 3 nonblank lines, medium proofs contain 4--10 lines, and long proofs contain more than 10 lines.
            We fixed an earlier implementation error that used generated proof length and recalculated the results from the audited raw data.
            For each model, we sum the generated tokens, generation time, and estimated \gls{gpu} time across all 100 attempts.
            We then divide each total by the number of verified proofs.
            Thus, the cost per verified proof includes both successful and unsuccessful attempts.

            \section{Findings}

            All checks listed in the pilot report passed.
            The run contained the planned 100 main theorems, 20 development theorems, six fixed model revisions, and 600 unique main model-theorem records.
            Every generation had a verification record, the protocol hashes matched, the leakage scan was clean, no accepted proof contained a forbidden construct, and all 21 model-theorem successes passed fresh-container reverification.
            Three of the six models produced at least one verified proof.
            The 21 accepted model-theorem results covered 15 distinct theorems, including 11 not solved by the fixed baseline.
            All model successes occurred in the short and medium reference-proof groups.
            Cost per observed success was defined for only three models.
            The sections below give the evidence behind each part of this summary, organized by research question.

            \subsection{Findings of $RQ_1$}

            The observed verified@1 rate ranged from 0\% to 12\%.
            Each model's two rates and their own paired interval are reported side by side, since the Wilson interval and the project-clustered bootstrap interval estimate two different quantities rather than offering two views of the same number.

            \begin{table}[htbp]
                \vspace{-10pt}
                \centering
                \scriptsize
                \small
                \setlength{\tabcolsep}{3.5pt}
                \caption{Verification rates and confidence intervals by model.}
                \label{tab:verification-rates}

                \begin{tabular}{lccc}
                    \toprule
                    \textbf{Model} &
                    \textbf{Verify} &
                    \textbf{Theorem Rate} &
                    \textbf{Project Rate} \\
                    \midrule
                    Llama
                    & 8/100
                    & 8.0\% [4.1\%, 15.0\%]
                    & 8.2\% [4.0\%, 12.6\%] \\

                    Qwen
                    & 0/100
                    & 0.0\% [0.0\%, 3.7\%]
                    & 0.0\% [0.0\%, 0.0\%] \\

                    Gemma
                    & 12/100
                    & 12.0\% [7.0\%, 19.8\%]
                    & 11.2\% [6.7\%, 15.4\%] \\

                    Mistral
                    & 0/100
                    & 0.0\% [0.0\%, 3.7\%]
                    & 0.0\% [0.0\%, 0.0\%] \\

                    GPT-OSS
                    & 0/100
                    & 0.0\% [0.0\%, 3.7\%]
                    & 0.0\% [0.0\%, 0.0\%] \\

                    DeepSeek
                    & 1/100
                    & 1.0\% [0.2\%, 5.4\%]
                    & 1.2\% [0.0\%, 3.6\%] \\
                    \bottomrule
                \end{tabular}
                \\[3pt]
                \raggedright \scriptsize Note: ``Verify'' denotes the number of verified proofs out of
                100 evaluated instances. Theorem Rate is weighted equally across theorems
                and uses Wilson 95\% confidence intervals. Project Rate is weighted equally
                across projects and uses cluster-bootstrap 95\% confidence intervals.
                \vspace{-5pt}
            \end{table}

            \textsc{Gemma} produced 12 verified proofs, \textsc{Llama} produced 8, and \textsc{DeepSeek} produced 1.
            \textsc{Qwen}, \textsc{Mistral}, and \textsc{GPT-OSS} produced no verified proofs under this one-attempt protocol.
            A zero observed count does not establish that a model's true success probability is exactly zero: the theorem-weighted Wilson interval still extends to 3.7\%.
            The project-weighted bootstrap collapses to [0\%, 0\%] for these three models specifically because it resamples from twelve observed project clusters that contain no success.
            We have explained in Methodology for why we treat that interval as a secondary sensitivity check rather than as evidence the true rate is zero.

            This pilot does not statistically test whether one model performs better than another.
            Although all six models attempt the same 100 theorems, we calculate each model’s confidence interval separately.
            A future study could compare two models by examining their results on each shared theorem and accounting for differences among projects.
            We do not compare confidence-interval overlap or use it to rank the models.

            \subsection{Findings of $RQ_2$}

            The 100 main theorems came from 12 projects.
            \textsc{Gemma} verified at least one theorem in 9 projects, \textsc{Llama} in 7, and \textsc{DeepSeek} in 1.
            Together, the models verified theorems in 10 of the 12 projects.
            None verified a theorem from projects \textit{hoare-tut} or \textit{reglang}.
            \textsc{Qwen}, \textsc{Mistral}, and \textsc{GPT-OSS} verified zero theorems in every project and every proof-length category, so the tables below show only the three models with at least one success, plus the fixed baseline.

            \begin{table}[htbp]
                \centering
                \setlength{\tabcolsep}{5pt}
                \caption{Number of verified theorems by project and model.}
                \label{tab:project-theorem-results}
                \begin{tabular}{lrrrrr}
                    \toprule
                    \textbf{Project} &
                    \textbf{Theorems} &
                    \textbf{Llama} &
                    \textbf{Gemma} &
                    \textbf{DeepSeek} &
                    \textbf{Baseline} \\
                    \midrule
                    buchberger  & 10 & 1 & 2 & 0 & 3 \\
                    compcert    & 10 & 2 & 2 & 0 & 0 \\
                    dblib       & 5  & 1 & 1 & 0 & 0 \\
                    ext-lib     & 10 & 0 & 1 & 0 & 3 \\
                    fourcolor   & 10 & 0 & 1 & 0 & 0 \\
                    hoare-tut   & 1  & 0 & 0 & 0 & 0 \\
                    huffman     & 10 & 1 & 1 & 0 & 0 \\
                    math-classes& 10 & 0 & 1 & 0 & 0 \\
                    poltac      & 7  & 1 & 1 & 0 & 0 \\
                    reglang     & 10 & 0 & 0 & 0 & 0 \\
                    zfc         & 7  & 1 & 0 & 1 & 0 \\
                    zorns-lemma & 10 & 1 & 2 & 0 & 1 \\
                    \bottomrule
                \end{tabular}
                \vspace{-10pt}
            \end{table}

            The reference-proof-length analysis was exploratory.
            Of the 100 theorems, 34 had short reference proofs, 43 had medium-length reference proofs, and 23 had long reference proofs.
            No model verified a theorem in the long category.
            This pattern is consistent with shorter reference proofs being easier under this protocol, but it is not enough to establish that proof length caused the difference.
            Reference-proof length is only an indirect measure of difficulty, and these groups were examined after generation.
            The fixed Coq automation baseline verified 7 of the 100 theorems.
            Its overlap with each of the successful models was as follows; \textsc{Qwen}, \textsc{Mistral}, and \textsc{GPT-OSS} contribute 0 to every column here by construction, since none produced a verified proof.

            \begin{table}[htbp]
                \vspace{-10pt}
                \centering
                \setlength{\tabcolsep}{11pt}
                \caption{Verification performance by problem length.}
                \label{tab:verification-by-length}
                \begin{tabular}{lccc}
                    \toprule
                    \textbf{Model} &
                    \textbf{Short (34)} &
                    \textbf{Medium (43)} &
                    \textbf{Long (23)} \\
                    \midrule
                    Llama
                    & 5/34 (14.7\%)
                    & 3/43 (7.0\%)
                    & 0/23 (0.0\%) \\

                    Gemma
                    & 8/34 (23.5\%)
                    & 4/43 (9.3\%)
                    & 0/23 (0.0\%) \\

                    DeepSeek
                    & 1/34 (2.9\%)
                    & 0/43 (0.0\%)
                    & 0/23 (0.0\%) \\
                    \bottomrule
                \end{tabular}
                \\[3pt]
                \raggedright \scriptsize Note: Each cell reports the number of verified instances,
                the total number of instances, and the corresponding verification
                rate in parentheses.
                \vspace{-5pt}
            \end{table}

            The six models produced 21 verified results across 15 distinct theorems because multiple models solved some of the same theorems.
            The baseline also solved 4 of these 15 theorems.
            The \glspl{llm} solved 11 theorems that the baseline could not solve, while the baseline solved 3 theorems that none of the \glspl{llm} could solve.
            Thus, the \glspl{llm} and baseline solved some different theorems.
            However, this result applies only to our limited three-tactic baseline and does not show that the LLMs outperform stronger or project-specific Coq automation.

            \begin{table}[htbp]
                \vspace{-10pt}
                \centering
                \scriptsize
                \setlength{\tabcolsep}{4pt}
                \caption{Comparison of model and baseline verification outcomes.}
                \label{tab:model-baseline-comparison}
                \begin{tabular}{lrrrr}
                    \toprule
                    \textbf{Model} &
                    \textbf{Both} &
                    \textbf{Model Only} &
                    \textbf{Baseline Only} &
                    \textbf{Neither} \\
                    \midrule
                    Llama 3.3 70B Instruct & 2 & 6 & 5 & 87 \\
                    Gemma 4 31B IT & 4 & 8 & 3 & 85 \\
                    DeepSeek Coder V2 Lite Instruct & 0 & 1 & 7 & 92 \\
                    \bottomrule
                \end{tabular}
                \\[3pt]
                \raggedright \scriptsize
                Note: ``Both'' indicates that both the model and the baseline verified
                the instance; ``Model Only'' indicates that only the model verified it;
                ``Baseline Only'' indicates that only the baseline verified it; and
                ``Neither'' indicates that neither verified the instance.
                \vspace{-15pt}
            \end{table}

            \subsection{Findings of $RQ_3$}

            We report each model’s total resource use across all 100 attempts.
            For models that verify at least one proof, we also divide the total resource use by the number of verified proofs.
            We cannot calculate this value for models with no verified proofs, and it may be unstable when a model has few successes.
            Therefore, we report total resource use and resource use per verified proof in separate tables.

            \subsubsection{Total resources across all 100 main attempts, all six models}

            \begin{table}[htbp]
                \vspace{-10pt}
                \centering
                \scriptsize
                \setlength{\tabcolsep}{5pt}
                \caption{Total output tokens, wall-clock time, and GPU-hours.}
                \label{tab:resource-usage}
                \begin{tabular}{lrrr}
                    \toprule
                    \textbf{Model} &
                    \textbf{Output Tokens} &
                    \textbf{Wall Time} &
                    \textbf{GPU Hours} \\
                    \midrule
                    Llama 3.3 70B Instruct
                    & 5{,}930
                    & 119.5\,s
                    & 0.133 \\

                    Qwen 3.5 35B-A3B
                    & 407{,}025
                    & 261.7\,s
                    & 0.300 \\

                    Gemma 4 31B IT
                    & 52{,}202
                    & 206.7\,s
                    & 0.233 \\

                    Mistral Small 3.1 24B Instruct
                    & 118{,}996
                    & 139.9\,s
                    & 0.150 \\

                    GPT-OSS 20B
                    & 307{,}349
                    & 103.6\,s
                    & 0.117 \\

                    DeepSeek Coder V2 Lite Instruct
                    & 36{,}193
                    & 178.0\,s
                    & 0.200 \\
                    \bottomrule
                \end{tabular}
                \\[3pt]
                \raggedright \scriptsize
                Note: Output Tokens denotes the total number of generated tokens;
                Wall Time denotes the total elapsed inference time; and GPU Hours
                denotes the total GPU usage in hours.
                \vspace{-5pt}
            \end{table}

            Models with no verified proofs still used substantial resources.
            For example, Qwen generated 407,025 tokens, while GPT-OSS generated 307,349 tokens.
            However, we cannot calculate tokens per verified proof for these models because they had no verified proofs.
            High token totals may also reflect long, repeated, or truncated responses rather than useful proof generation.

            \subsubsection{Total cost per verified proof, models with at least one success}

            \begin{table}[htbp]
                \vspace{-10pt}
                \centering
                \scriptsize
                \setlength{\tabcolsep}{3.25pt}
                \caption{Resource usage per verified instance.}
                \label{tab:resource-per-verified}
                \begin{tabular}{lrrrr}
                    \toprule
                    \textbf{Model} &
                    \textbf{Verif.} &
                    \textbf{Tok./Verif.} &
                    \textbf{Time/Verif.} &
                    \textbf{GPU h/Verif.} \\
                    \midrule
                    Llama 3.3 70B Instruct
                    & 8
                    & 741.2
                    & 14.9\,s
                    & 0.0167 \\

                    Gemma 4 31B IT
                    & 12
                    & 4{,}350.2
                    & 17.2\,s
                    & 0.0194 \\

                    DeepSeek Coder V2 Lite Instruct
                    & 1
                    & 36{,}193.0
                    & 178.0\,s
                    & 0.2000 \\
                    \bottomrule
                \end{tabular}
                \\[3pt]
                \raggedright \scriptsize
                Note: ``Verif.'' denotes the number of verified instances.
                ``Tok./Verif.'' denotes the average number of output tokens per
                verified instance. ``Time/Verif.'' denotes the average wall-clock
                time per verified instance, and ``GPU h/Verif.'' denotes the average
                GPU usage in hours per verified instance.
                \vspace{-5pt}
            \end{table}


            \textsc{Llama} used the fewest resources per verified proof, followed by \textsc{Gemma}.
            \textsc{DeepSeek}’s value is less reliable because it verified only one proof, so its total cost across 100 attempts is divided by one.
            If \textsc{DeepSeek} had verified one more proof with the same total resource use, its cost per verified proof would be cut in half.
            These results describe this specific run on four \glspl{gpu} and should not be treated as a general efficiency ranking.

            \section{Discussion}

            Under the strict one-attempt setting, verified proofs were rare.
            Only three of the six models produced a verified proof, and the highest success rate was 12\%.
            However, these results show that open-weight models can generate complete proofs that pass all verification checks in real Coq projects.

            The failures show that models struggled with more than logical reasoning.
            \textsc{Qwen} produced 98 truncated responses.
            \textsc{GPT-OSS} produced 59 truncated responses and 41 syntax errors.
            \textsc{Mistral} produced 60 syntax errors, 28 truncated responses, and 9 empty responses, while \textsc{DeepSeek} produced 93 syntax errors.
            \textsc{Llama} and \textsc{Gemma} also had many project-environment errors.
            Therefore, verified@1 measures the complete proof-generation process, including correct syntax, output format, and compatibility with project libraries.

            The LLMs and the fixed automation baseline solved different theorems.
            The LLMs solved 11 theorems that the baseline could not solve, while the baseline solved 3 that none of the LLMs could solve.
            This result suggests that LLMs and basic automation may work well together.
            However, it does not show that LLMs outperform Coq automation because we did not test stronger or project-specific automation.

            Total resource use remains meaningful for every model, including models with no verified proofs.
            For example, \textsc{Qwen} generated 407,025 tokens and used 0.300 GPU hours without verifying any proof.
            In contrast, resource use per verified proof is unstable when there are few successes.
            \textsc{DeepSeek} verified only one proof, so one additional success with the same total cost would reduce its cost per verified proof by half.
            So, these values should not be treated as a general efficiency ranking.

            Success also varied across projects and proof lengths.
            \textsc{Gemma} verified proofs from nine projects, \textsc{Llama} from seven, and \textsc{DeepSeek} only from \texttt{zfc} project.
            Every verified theorem had a short or medium reference proof.
            However, the small number of successes and the exploratory proof-length analysis do not establish which project or theorem features cause success.

            \section{Threats to Validity}

            This study evaluates six fixed model versions on 100 theorems using one prompt and one attempt per theorem.
            The results apply only to this setup and do not represent performance on all formal proofs.
            Different prompts, repeated attempts, retrieval, repair, tool access, or additional context could change the results.

            The 100 theorems come from 12 projects, with no more than 10 from each project.
            This limit increases project diversity but makes the sample different from the full CoqStoq dataset.
            Because some projects and proof-length groups contain few theorems, their results have substantial uncertainty.
            The project-level confidence intervals are also limited by the small number of projects.

            Reference-proof length is only a rough measure of difficulty.
            A long human proof may have a short automated solution, while a short proof may require specialized knowledge.
            We therefore treat the proof-length analysis as exploratory.

            The resource measurements depend on the hardware and software used in this study.
            GPU use was recorded every 15 seconds, so the reported GPU time may be slightly higher or lower than the actual value.

            Many \textsc{Llama} and \textsc{Gemma} failures were labeled \texttt{project\_environment\_error}.
            This label indicates difficulty producing a proof that works with the project’s libraries, notation, or configuration.
            Therefore, these failures do not measure logical reasoning alone.

            The automation baseline uses only \texttt{auto}, \texttt{eauto}, and \texttt{intuition auto}.
            It provides a basic comparison but does not represent stronger, specialized, or expert-designed Coq automation.
            Claims about LLM-only successes apply only to this fixed baseline.

            Finally, this pilot does not statistically test whether one model outperforms another.
            A future study could compare the models theorem by theorem while accounting for differences among projects.
            The current results are descriptive and should not be treated as a universal model ranking.

            \section{Conclusion}

            Under a strict one-shot, no-repair protocol, open-weight LLMs occasionally produced fully kernel-verified Coq proofs in real projects, including proofs a minimal fixed tactic baseline could not find.
            Their low yield and frequent formatting, truncation, and project-interface failures, however, make them unreliable as standalone proof automation under this protocol.
            Three of the six models (\textsc{Qwen}, \textsc{Mistral}, and \textsc{GPT-OSS}) never produced a single verified proof, and success for the other three was concentrated in short and medium-length reference proofs.
            Models that verified no proofs still used substantial tokens and GPU time.
            For successful models, the cost per verified proof is uncertain because it is based on only a few successes—and on just one success for \textsc{DeepSeek}.

            Future studies should compare LLMs with stronger and project-specific Coq automation.
            They should separate syntax and project-environment failures from failures in logical reasoning.
            They should also test whether repair, retrieval, and repeated attempts improve performance.
            Because every model evaluates the same theorems, future model comparisons should compare their results theorem by theorem.
            These studies should continue using Coq kernel verification and independent reverification to ensure that every accepted proof is valid.

            \bibliographystyle{IEEEtran}
            \bibliography{references}

        \end{document}